\documentclass[12pt]{article}
\usepackage{graphicx}
\usepackage{amsmath}
\usepackage{indentfirst}
\usepackage{float} 
\usepackage{booktabs}
\usepackage{graphicx}
\usepackage{times} 
\usepackage{caption}
\usepackage{geometry}

\title{{Regularization method in the variable selection for logistic regression on BRFSS data}}
\author{
  Jinbo Niu\\[0.5em]
  \small University of Pennsylvania, School of Social Policy \& Practice\\
  \small {jn1014@upenn.edu}
}

\date{} 

\begin{document}

\maketitle
\section*{Abstract}

Stroke remains a leading cause of death and disability worldwide, yet effective prediction of stroke risk using large-scale population data remains challenging due to data imbalance and high-dimensional features. In this study, we develop and evaluate regularized logistic regression models for stroke prediction using data from the 2022 Behavioral Risk Factor Surveillance System (BRFSS), comprising 445,132 U.S. adult respondents and 328 health-related variables. To address data imbalance, we apply several resampling techniques including oversampling, undersampling, class weighting, and the Synthetic Minority Oversampling Technique (SMOTE). We further employ Lasso, Elastic Net, and Group Lasso regularization methods to perform feature selection and dimensionality reduction. Model performance is assessed using ROC-AUC, sensitivity, and specificity metrics. Among all methods, the Lasso-based model achieved the highest predictive performance (AUC = 0.761), while the Group Lasso method identified a compact set of key predictors—Age, Heart Disease, Physical Health, and Dental Health. These findings demonstrate the potential of regularized regression techniques for interpretable and efficient prediction of stroke risk from large-scale behavioral health data.

\section*{Introduction}

Stroke remains a major global health concern, ranking as the fifth leading cause of death in the United States and a significant contributor to severe long-term disability worldwide\cite{Ahmed2025}\cite{Katan2018} . Despite advances in medical care, the prompt identification of stroke symptoms in acute settings continues to pose challenges, often resulting in missed opportunities for timely intervention \cite{Hsia2011}\cite{Boehme2017}. As a result, stroke persists as a threat to public health, which requires improved strategies for prevention, detection, and treatment.

The complex etiology of stroke involves multiple risk factors, including hypertension, smoking, diabetes, and obesity \cite{Kono2023}\cite{SHANG2025108313}\cite{Ng2020_LifestyleRiskFactors}. Understanding these risk factors and their interplay is crucial for developing effective prevention and management strategies\cite{Bushnell2024}. Additionally, socioeconomic factors and access to healthcare have been shown to influence stroke outcomes, highlighting the need for a comprehensive approach to stroke research and care\cite{Pantoja-Ruiz2025}.

To address these challenges, large-scale epidemiological studies and surveillance systems play a vital role in monitoring stroke trends and risk factors across populations\cite{Goff2024} One such initiative is the Behavioral Risk Factor Surveillance System (BRFSS), an annual telephone survey conducted in the United States\cite{CDC_BRFSS}. The BRFSS aims to identify risk factors and track emerging health trends among adults, encompassing a wide range of topics including diet, physical activity, HIV/AIDS status, tobacco use, immunizations, existing health conditions, access to healthcare, and alcohol consumption\cite{CDC_BRFSS}.

The data applied in this research were collected from all 50 states, the District of Columbia, Guam, Puerto Rico, and the US Virgin Islands by conducting landline telephone surveys and cell phone-based surveys in 2022. Disproportionate stratified sampling (DSS) has been used for the landline sample, and cell phone respondents are randomly selected, with each respondent having an equal probability of selection. The dataset we are working with contains 328 variables and a total of 445,132 observations in 2022. Missing values are represented by "NA". The data was retrieved from the CDC official website and it is originally in the SAS transport format.

The data consists of survey responses from 445,132 U.S. adults aged 18 and over. It is based on a large stratified random sample. Potential biases including non-response, incomplete interviews, missing values, and convenience bias. To alleviate the bias, some responses with blank, refused to answer, and not sure will be removed.

From the 328 variables included in the dataset, 20 are considered in the analysis. The variables are renamed and also recoded as below for further analysis:

\begin{description}
  \item[Age:] Age of the patient
  \item[Smoke:] Smoke at least 100 cigarettes (0: No, 1: Yes)
  \item[Drinking:] Average alcoholic beverages per day over the past 30 days
  \item[BMI:] Computed body mass index (BMI)
  \item[Mental:] Number of days mental health not good
  \item[Physical:] Number of days physical health not good
  \item[COPD:] Ever diagnosed with COPD (0: No, 1: Yes)
  \item[Education:] 1: Elementary, 2: Some high school, 3: High school graduate, 4: Some college, 5: College or more
  \item[Veteran:] Veteran status (0: No, 1: Yes)
  \item[Dental:] Teeth removed (0: 0 teeth, 1: 1--5 teeth, 2: 6+ teeth)
  \item[Sleep:] Hours of sleep per day (1--24)
  \item[Insurance:] Have any health insurance (0: No, 1: Yes) 
  \item[Male:] Sex (0: Female, 1: Male)
  \item[Stroke:] Ever diagnosed with stroke (0: No, 1: Yes)
  \item[Heart:] Ever had a heart attack (0: No, 1: Yes)
  \item[Visual:] Blind or visual impairment (0: No, 1: Yes)
  \item[Hearing:] Serious difficulty hearing or deaf (0: No, 1: Yes)
  \item[Drug:] Injected any non-prescribed drug (0: No, 1: Yes)
  \item[Type I:] Type I diabetes (0: No, 1: Yes)
  \item[Type II:] Type II diabetes (0: No, 1: Yes)
\end{description}

\section*{Material \& Methods}

The 2022 Behavioral Risk Factor Surveillance System (BRFSS) data will serve as the foundation for our data analysis, with R chosen as the programming language for both data wrangling and modeling tasks. Initially, the data will be transformed from the SAS format to an R-compatible file format, enabling seamless integration into our analytical workflow. Subsequently, the modeling and analysis phases will be conducted exclusively within the R environment.

Our predictive model will include a diverse set of variables, including gender, education level, smoking habits, alcohol consumption, sleep patterns, diabetes status, drug usage, BMI, age, history of heart attack, visual and auditory impairments, veteran status, physical and mental health indicators, presence of depression disorder, oral health status, and chronic obstructive pulmonary disease (COPD). Prior to model construction, we will address the challenge of imbalanced data, a common issue in stroke prediction tasks, where the majority of patients have not experienced a stroke. To mitigate this imbalance, we will employ various techniques such as class weights adjustment, synthetic sample generation, over-sampling, and under-sampling.

The performance of each imbalanced data handling method will be assessed using Receiver Operating Characteristic (ROC) analysis, enabling us to identify the most effective approach for further model development. Logistic regression, elastic net regularization, and lasso methods, including both standard lasso and group lasso techniques, will be employed as our learning algorithms.

Initially, a logistic regression model will be built, incorporating all variables with a probability cutoff of 0.5. Then regularization techniques such as lasso, elastic net, and group lasso logistic regression will be applied, with the optimal regularization parameter (lambda) determined through 10-fold cross-validation.

Following regression analysis, feature selection will be performed to identify the most informative variable sets. New logistic regression models will be constructed based on these selected features, allowing for a comparative evaluation of model accuracies between the initial and refined models. Variables demonstrating consistent predictive power across these models will be deemed as the most critical risk factors.

To avoid the issue of overfitting, 10-fold cross-validation will be employed throughout our modeling process, ensuring the robustness and generalizability of our predictive models.

\section*{Results}

\subsection*{Descriptive statistics results}
\subsubsection*{The Outcome variable}
\begin{figure}[H]
    \centering
    \includegraphics[width= .75\linewidth]{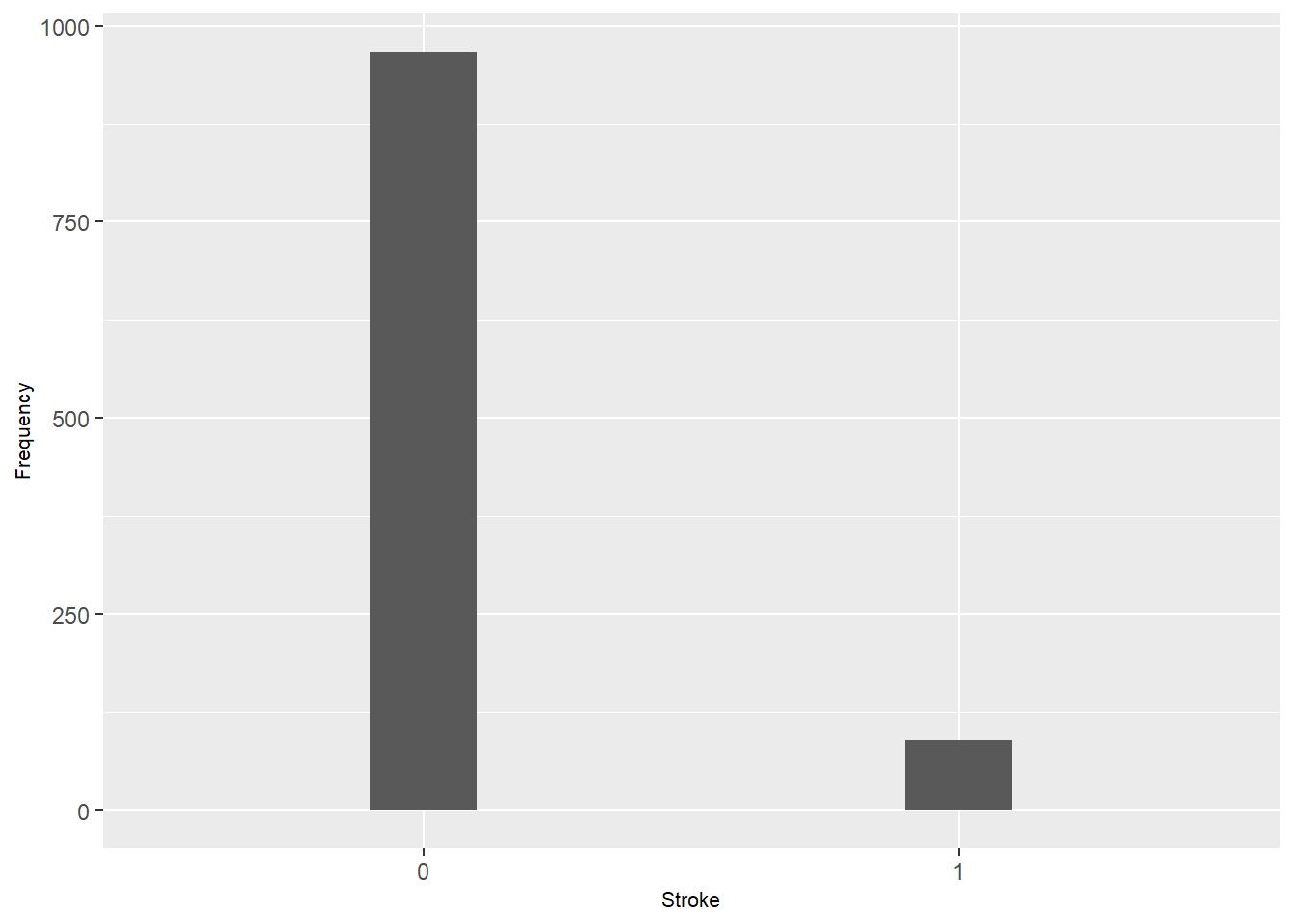}
    \caption{Distribution of the outcome variable}
    \label{fig:outcome}
\end{figure}

According to the bar chart representing the distribution of the outcome variable, it becomes evident that the dataset exhibits a notable imbalance, with a relatively small proportion of individuals experiencing stroke. This disparity in distribution is understandable, given the diverse age range of our study population and stroke occurrence may be less prevalent among certain age cohorts.

Recognizing the importance of addressing this class imbalance to ensure the robustness of our predictive model, we will employ various strategies such as oversampling, under-sampling, and the Synthetic Minority Over-sampling Technique (SMOTE)\cite{Wu2020}\cite{Joloudari2023}. These techniques aim to rectify the imbalance by either augmenting the minority class instances, reducing the dominance of the majority class, or generating synthetic samples to balance the dataset \cite{Vargas2023_ImbalancedData}. By implementing these actions, we aim to enhance the model's ability to accurately predict stroke occurrences across all demographic groups, thereby improving its overall performance and reliability\cite{Bushnell2024}.

\subsubsection*{Categorical variables}

The series of bar plots are the distributions of categorical variables. Generally most variables have even distributions. For some variables like diabetes and heart attack, the distribution could be as uneven as the outcome variable but the distributions are still reasonable as most people are not likely to have those diseases.

Also, we noticed that the patients not having type I diabetes are all having type II diabetes. So in this case, we will only take the Type I variable into the model.

\begin{figure}[H]
    \centering
    \includegraphics[width=1\linewidth]{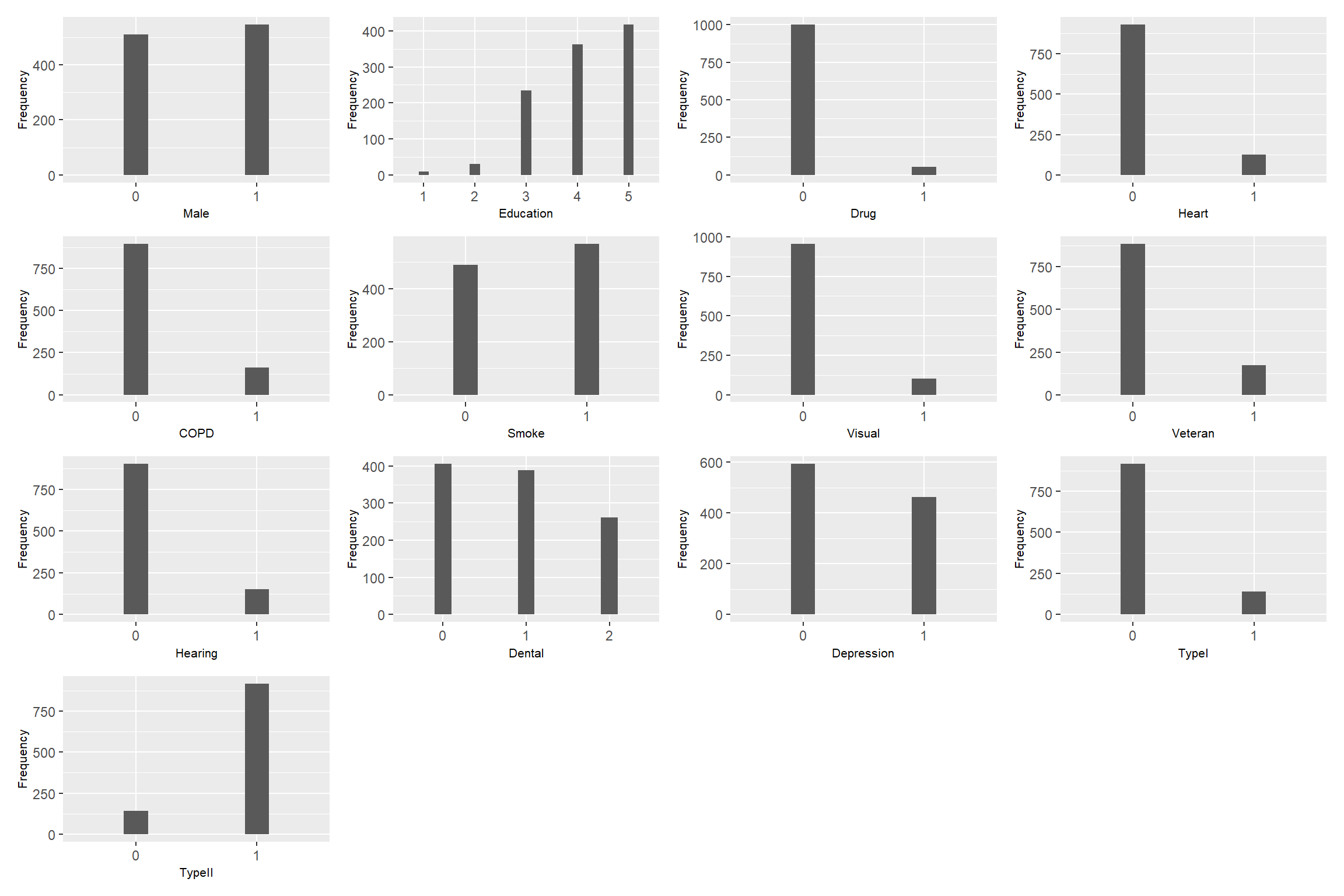}
    \caption{Distributions of categorical variables}
    \label{fig:categorical}
\end{figure}

\subsubsection*{Continuous variables}

Box plots are utilized to visually represent the distribution of continuous variables. To ensure uniformity across variables, a log10 transformation is implemented. It is notable that the majority of variables exhibit outliers; however, given their behavioral nature, these outliers are deemed reasonable and thus are retained rather than removed.

Provided below is a comprehensive table detailing the descriptive statistics of the continuous variables:

\begin{table}[H]
\centering
\begin{tabular}{lcccccccc}
\hline
\textbf{name} & \textbf{n} & \textbf{mean} & \textbf{sd} & \textbf{median} & \textbf{max} & \textbf{min} & \textbf{skew} & \textbf{se} \\
\hline
Sleep & 1057 & 6.80 & 1.95 & 7 & 23 & 1 & 1.42 & 0.06 \\
BMI & 1057 & 3264.94 & 736.80 & 3165 & 7586 & 1414 & 0.85 & 22.66 \\
Age & 1057 & 58.41 & 13.46 & 60 & 80 & 18 & -0.50 & 0.41 \\
Drinking & 1057 & 2.30 & 3.42 & 2 & 72 & 1 & 11.38 & 0.11 \\
Mental & 1057 & 11.83 & 10.38 & 8 & 30 & 1 & 0.75 & 0.32 \\
Physical & 1057 & 9.11 & 11.30 & 3 & 30 & 0 & 0.98 & 0.35 \\
\hline
\end{tabular}
\caption{Summary statistics of continuous variables}
\label{tab:summary}
\end{table}

\begin{figure}[H]
    \centering
    \includegraphics[width=1\linewidth]{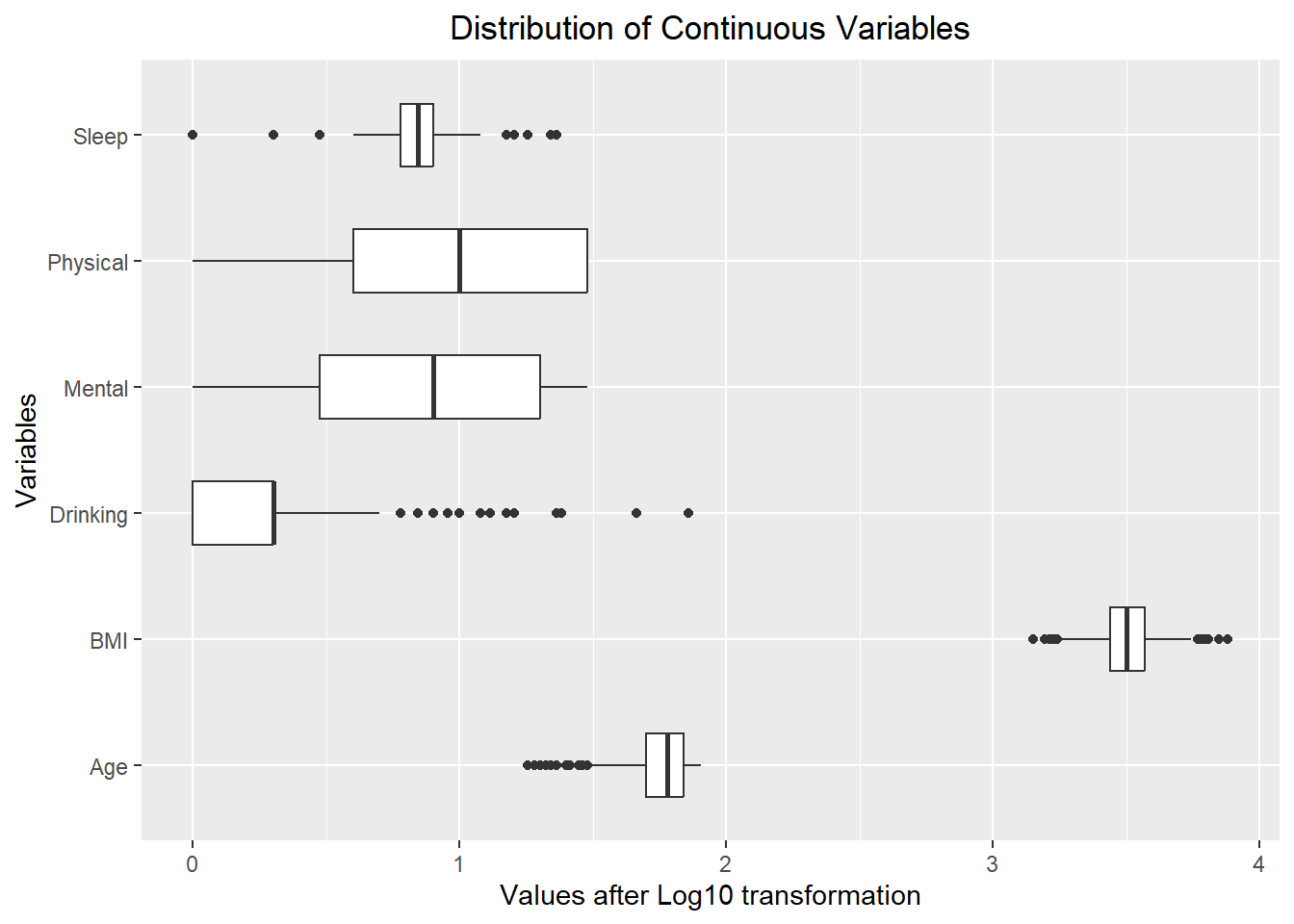}
    \caption{Distribution of continuous variables}
    \label{fig:continuous}
\end{figure}

\subsubsection*{Data Sampling results}

To address the challenge of class imbalance, four methods—class weights, oversampling, under-sampling, and Synthetic Minority Over-sampling Technique (SMOTE)—were employed. Analysis of the results presented in the table reveals that the SMOTE method yields the most favorable performance, achieving the highest AUC among the evaluated methods. Although the Class Weight method demonstrates a competitive AUC value of 0.7396, its low sensitivity of approximately 0.067 raises concerns regarding its effectiveness in correctly identifying positive cases. As such, despite its comparable AUC performance, the Class Weight method is deemed less suitable for addressing imbalanced data due to its poor sensitivity. Consequently, the SMOTE method emerges as the preferred approach for handling class imbalance in subsequent analyses.

\begin{table}[H]
\centering
\begin{tabular}{|l|c|c|c|}
\hline
Name & AUC & Sens & Spec \\
\hline
SMOTE & 0.739851 & 0.6 & 0.725773 \\
Over sampling & 0.728938 & 0.633333 & 0.699173 \\
Under sampling & 0.717773 & 0.6 & 0.677298 \\
Class weight & 0.7396 & 0.066667 & 0.992751 \\
\hline
\end{tabular}
\caption{Data Sampling results}
\end{table}

\subsection*{Modeling results}
\subsubsection*{grouping results}

Five grouping policies were employed in this study. The first policy involves grouping variables based on their data type, resulting in two distinct groups: continuous variables and categorical variables. The second policy categorizes variables into three broad categories: Demographic, Behavior, and Health condition. The third policy further subdivides variables into four categories: Demographic, Behavior, Physical, and Mental Health. The fourth policy expands upon the third, adding Dental as a separate category. Lastly, the fifth policy treats each variable as an individual group. The outcomes of these grouping policies are detailed in Table 3.

\begin{table}[H]
  \centering
  \caption{Grouped variables}
  \begin{tabular}{|l|c|c|c|c|c|}
    \hline
    Variables   & group 1 & group 2 & group 3 & group 4 & group 5 \\
    \hline
    Male        & 1               & 1               & 1               & 1               & 1               \\
    Education   & 2               & 1               & 1               & 1               & 2               \\
    Sleep       & 2               & 2               & 2               & 2               & 3               \\
    Drug        & 1               & 2               & 2               & 2               & 4               \\
    BMI         & 2               & 1               & 1               & 1               & 5               \\
    Age         & 2               & 1               & 1               & 1               & 6               \\
    Heart       & 1               & 3               & 3               & 3               & 7               \\
    Drinking    & 2               & 2               & 2               & 2               & 8               \\
    Mental      & 2               & 3               & 4               & 4               & 9               \\
    COPD        & 1               & 3               & 3               & 3               & 10              \\
    Smoke       & 1               & 2               & 2               & 2               & 11              \\
    Visual      & 1               & 2               & 2               & 2               & 12              \\
    Veteran     & 1               & 3               & 3               & 3               & 13              \\
    Hearing     & 1               & 1               & 1               & 1               & 14              \\
    Physical    & 2               & 3               & 3               & 3               & 15              \\
    Dental      & 2               & 3               & 3               & 5               & 16              \\
    Depression  & 1               & 3               & 4               & 4               & 17              \\
    TypeI       & 1               & 3               & 3               & 3               & 18              \\
    \hline
  \end{tabular}
  \label{tab:variables_grouped}
\end{table}

\subsubsection*{variable selection results}

Utilizing various regularization techniques such as lasso, elastic net, and group lasso with distinct groupings, certain variables have had their coefficients effectively shrunk to zero. Consequently, variables associated with zero coefficients are deemed redundant and will be eliminated from further analysis, while those retaining non-zero coefficients will be retained for model refinement. Implementation of group lasso methods was facilitated through the gglasso package; however, it is noteworthy that this package has the propensity to shrink all variables to zero. To mitigate this potential limitation, alternative groupings may be explored, prioritizing groups with the least number of non-zero coefficients to ensure the retention of informative variables.

\begin{table}[H]
    \centering
    \caption{Variables removed by non-group lasso methods}
    \begin{tabular}{|c|c|}
    \hline
    \textbf{Model} & \textbf{Variables removed} \\ \hline
    Stepwise & Education, BMI, Hearing \\ \hline
    Lasso & Sleep, BMI, Drinking, Smoke, Hearing, TypeI \\ \hline
    Elasticnet & BMI, Drinking, Hearing \\ \hline
    \end{tabular}
    \label{tab:variables_removed}
\end{table}

\begin{table}[htbp]
\centering
\caption{Variable Selection Results of gorup lasso models}
\label{tab:variable_selection}
\resizebox{\linewidth}{!}{%
\begin{tabular}{lcccccccccc}
\hline
\textbf{Variables} & \textbf{group1} & \textbf{group2} & \textbf{group3} & \textbf{group4(1)} & \textbf{group4(2)} & \textbf{group5(1)} & \textbf{group5(2)} & \textbf{group5(3)} & \textbf{group5(4)} \\ \hline
Male &  &  &  &  &  &  &  &  &  \\
Education &  &  &  &  &  &  &  &  & * \\
Sleep &  &  &  &  &  &  &  &  &  \\
Drug &  &  &  &  &  &  &  &  &  \\
BMI &  &  &  &  &  &  &  &  &  \\
Age &  & * & * &  & * &  & * & * & * \\
Heart &  & * & * &  & * &  &  & * & * \\
Drinking &  & * & * &  &*  &  &  &  &  \\
Mental &  & * & * &  & * & &  &  & * \\
COPD &  & * & * &  & * &  &  &  &  \\
Smoke & * &  & * &  &  &  &  &  &  \\
Visual & * &  & * &  &  &  &  &  &  \\
Veteran & * &  & * &  &  &  &  &  &  \\
Hearing & * &  & * &  &  &  &  &  &  \\
Physical & * &  & * & &  & * & * & * & * \\
Dental & * &  & * & * & * &  & * & * & * \\
Depression & * &  &  &*  & * &  &  &  &  \\
TypeI & * &  &  &  &  &  &  &  &  \\ \hline
\end{tabular}%
}
\end{table}

\subsubsection*{Logistic Models with different variable selections}

In this section, the selected variables will be incorporated into the logistic regression model and compared with the original logistic regression model comprising all variables. The performance of each model will be assessed based on key metrics such as the Area Under the Curve (AUC), Sensitivity, and Specificity, as presented in Table 6. Upon examination, it becomes evident that the logistic model with the variables selected from the Lasso model exhibits the most favorable performance, boasting the highest AUC among all models tested. Following closely behind is the logistic model with variables from group 5(3), which achieves an AUC of 0.7488 and a sensitivity of 0.6111, showcasing minimal discrepancies compared to the Lasso model. Notably, the group 5(3) derived from group lasso achieves dimension reduction from 18 variables to 4, surpassing the Lasso model's reduction from 18 to 12 variables. Consequently, the variables from group 5(3) derived from group lasso emerges as the preferred choice for the final model selection.

Therefore, four variables—\textbf{Age, Heart, Physical, and Dental}— are selected and are considered as significant predictors  for future research endeavors. These variables exhibit substantial predictive power and warrant further investigation in subsequent studies.

\begin{table}[H]
\centering
\begin{tabular}{llll}
\hline
Logistic model with selected variables & ROC & Sens & Spec \\
\hline
Full Logistic Regression & 0.7311 & 0.5778 & 0.7220 \\
variables from Stepwise & 0.7464 & 0.6111 & 0.7280 \\
variables from Lasso & 0.7613 & 0.6667 & 0.7259 \\
variables from Elastic Net & 0.7503 & 0.6333 & 0.7116 \\
variables from Group 1 & 0.6502 & 0.5556 & 0.6371 \\
variables from Group 2 & 0.7124 & 0.5333 & 0.7672 \\
variables from Group 3 & 0.7341 & 0.5889 & 0.7270 \\
variables from Group 4(1) & 0.6483 & 0.6222 & 0.5957 \\
variables from Group 4(2) & 0.7289 & 0.6222 & 0.7352 \\
variables from Group5(1) & 0.6465 & 0.4889 & 0.7115 \\
variables from Group5(2) & 0.6991 & 0.6333 & 0.6412 \\
variables from Group5(3) & 0.7488 & 0.6111 & 0.7311 \\
variables from Group5(4) & 0.7450 & 0.6000 & 0.7363 \\
\hline
\end{tabular}
\caption{Model performance based on different variable selections}
\label{tab:model_performance}
\end{table}

\section*{Discussion}

\subsection*{Modeling results and advantages}

In this study, we employed regularization methods with a particular focus on the group lasso technique, to effectively reduce dimensions and filter variables most pertinent to stroke. Regularization methods offer a powerful means to address the challenges of high-dimensional data by imposing penalties on model coefficients, thereby encouraging sparsity and feature selection\cite{Yokochi2023}. The group lasso, in particular, extends the benefits of traditional lasso regularization by incorporating group structures within the data, allowing for the simultaneous selection of entire groups of correlated variables\cite{Klosa2020_Seagull}. This approach not only helps identify key predictors associated with Stroke but also facilitates the interpretation of model results.

Compared to Principal Component Analysis (PCA), which operates solely on continuous variables and may obscure the interpretability of the underlying data, regularization methods offer a more flexible framework that can accommodate both continuous and categorical predictors\cite{Tutz2016}\cite{Nowakowski2023}. Moreover, while PCA may struggle with non-linear relationships and interactions among variables\cite{PortnovaFahreeva2020_HandKinematics}, regularization methods, such as the group lasso, are better equipped to capture complex patterns in the data, leading to more robust and interpretable models\cite{Friedrich2023_RegularizationReview}.

Despite the reduction in the number of variables retained by the group lasso method, our logistic models constructed with these selected variables consistently demonstrated strong performance in predicting stroke outcomes. This highlights the efficiency of the group lasso in identifying a compact subset of informative predictors while maintaining predictive accuracy\cite{Liu2020}. By prioritizing relevant variables and discarding noise, the group lasso effectively balances model complexity and performance, offering a practical solution for dimension reduction in predictive modeling tasks\cite{7887916}.

\subsection*{Limitations}

In terms of methodology, the group lasso method offers the advantage of preserving the correlation of features within groups and handling both continuous and categorical variables. However, defining these groups can be subjective, potentially impacting model performance and interpretability\cite{WANG20085277}. Moreover, the sensitivity of Group Lasso to outliers, particularly those within group clusters, can further affect model performance\cite{Jiang2020_PenalizedLADLasso}. Given the presence of outliers that cannot be removed, the efficacy of the group lasso model may be compromised.

Furthermore, when dealing with health status surveillance data, relying solely on self-reported prevalence rates may introduce bias, as respondents may lack awareness of their true risk status \cite{Rosenman2011_SelfReportBias}. Thus, to obtain more precise estimates, it may be necessary to supplement self-reported data with laboratory tests\cite{StClair2017_SelfReportsClaims}.

Additionally, the gglasso package may exhibit a tendency to shrink all variables to zero, necessitating a thorough examination of the coefficient matrix and individual group selections with less optimal lambda values\cite{Yang2015_GMD}. In such cases, alternative tools may be explored to complement the group lasso approach.

\section*{Conclusion}
This study demonstrates the effectiveness of regularization-based logistic regression models for stroke risk prediction using large-scale behavioral health survey data. By systematically evaluating Lasso, Elastic Net, and Group Lasso methods under various data balancing strategies, we show that regularization can achieve competitive predictive performance while maintaining model interpretability. The integration of SMOTE with Lasso-based models yielded the best trade-off between sensitivity and specificity, highlighting the importance of addressing data imbalance in health-related machine learning tasks.

From a methodological perspective, our findings emphasize that group-structured penalization methods can serve as a powerful and interpretable alternative to black-box models in population-level health prediction. The selected key predictors—Age, Heart Disease, Physical Health, and Dental Health—reflect consistent behavioral and physiological relevance, demonstrating that interpretable models can still provide high performance on large-scale heterogeneous data.

Future work will extend these models to nonlinear and ensemble approaches such as gradient boosting and neural networks, and explore domain adaptation for cross-year BRFSS data. Additionally, integrating explainability techniques (e.g., SHAP or LIME) will further enhance transparency and clinical trust in machine learning–driven public health modeling.

\begingroup

\bibliographystyle{unsrt}
\bibliography{main}
\endgroup
\end{document}